\documentstyle[12pt,epsf]{article}

\newlength{\dinwidth}
\newlength{\dinmargin}
\def\lapproxeq{\lower .7ex\hbox{$\;\stackrel{\textstyle
<}{\sim}\;$}}
\def\gapproxeq{\lower .7ex\hbox{$\;\stackrel{\textstyle
>}{\sim}\;$}}
\def\funp{{I\!\!P}}
\def\be{\begin{equation}}
\def\ee{\end{equation}}
\def\bea{\begin{eqnarray}}
\def\eea{\end{eqnarray}}
\makeatletter
\def\fmslash{\@ifnextchar[{\fmsl@sh}{\fmsl@sh[0mu]}}
\def\fmsl@sh[#1]#2{%
\mathchoice
{\@fmsl@sh\displaystyle{#1}{#2}}%
{\@fmsl@sh\textstyle{#1}{#2}}%
{\@fmsl@sh\scriptstyle{#1}{#2}}%
{\@fmsl@sh\scriptscriptstyle{#1}{#2}}}
\def\@fmsl@sh#1#2#3{\m@th\ooalign{$\hfil#1\mkern#2/\hfil$\crcr$#1
#3$}}
\makeatother

\begin{document}
\titlepage
\begin{flushright}
DTP/97/06 \\
hep-ph/9701366 \\
January 1997
\end{flushright}

\begin{center}
\vspace*{2cm}
{\Large \bf Diffractive electroproduction of \\[2mm]

$\rho$ meson excitations} \\

\vspace*{1cm}
A.\ D.\ Martin, M.\ G.\ Ryskin\footnote{Permanent address: 
Laboratory of Theoretical Nuclear Physics, St.~Petersburg Nuclear
Physics Institute, Gatchina, St.~Petersburg 188350, Russia} and
T.\ Teubner \\

\vspace*{0.5cm}
Department of Physics, University of Durham, Durham, DH1 3LE, UK.
\end{center}

\vspace*{3cm}
\begin{abstract}
We use perturbative QCD and parton-hadron duality to estimate the
cross sections for the diffractive electroproduction of
$\rho^\prime (1^-)$ and $\rho (3^-)$ resonances which occur in
the 1.3--1.8 GeV mass interval.  We present the cross sections
and the ratios $\sigma_L/\sigma_T$ as a function of $Q^2$.  We
compare the predictions with those for the diffractive
electroproduction of $\rho$ mesons.  We show how such diffractive
electroproduction measurements at HERA can probe features of the
perturbative QCD \lq Pomeron'.
\end{abstract}

\newpage

The experiments \cite{RHO} at HERA are measuring the diffractive
electroproduction processes $\gamma^* p \rightarrow (2 \pi) p$
and $\gamma^* p \rightarrow (4 \pi) p$ as a function of invariant
mass $M$ of the pionic system, for different intervals of $Q^2$,
the virtuality of the photon, and of $W$, the centre-of-mass
energy of the $\gamma^* p$ system.  As these are quasi-elastic
processes we would expect that at high energy the pionic system
will dominantly have spin-parity $J^P = 1^-$.  Indeed a strong
$\rho$ meson resonant peak is observed \cite{RHO}.  In a
previous
paper \cite{MRT} we presented a QCD model which reproduced the
observed $\rho$ electroproduction cross section and, in
particular, described the $Q^2$ behaviour of the cross section
ratio $\sigma_L/\sigma_T$ for $\rho$ meson production in
longitudinally and transversely polarised states.  Here we study
the rate of $\rho^\prime (1^-)$ resonance production in the
higher mass interval, 1.3--1.8 GeV.  Moreover we are also able to
estimate the cross section for the diffractive production of
the $\rho (3^-)$ resonance at HERA, the $\rho$ orbital excitation
which also occurs in the above mass interval.  Given sufficient
data, the comparison of $\rho^\prime (1^-)$ and $\rho (3^-)$
production would provide a unique opportunity to study how the
QCD \lq Pomeron' distorts the initial state.

\begin{figure}[htb]
\begin{center}
\vskip -1mm
\leavevmode
\epsfxsize=17.cm
\epsffile[123 359 542 527]{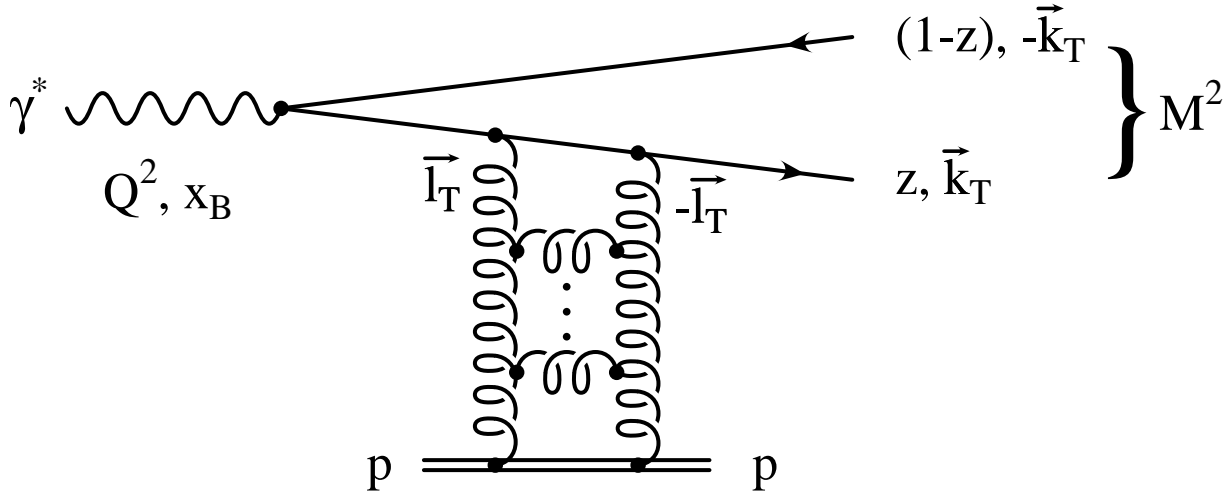}
\vskip -2mm
\caption[]{\label{fig1} {Diffractive open $q\overline{q}$ production in high
energy $\gamma^* p$ collisions.  The transverse momenta of the
outgoing quarks are $\pm \mbox{\boldmath $k$}_T$, and those of
the exchanged gluons are $\pm \mbox{\boldmath $\ell$}_T$.}} 
\end{center}
\end{figure}

A perturbative QCD description of diffractive $\rho$
electroproduction, $\gamma^* p \rightarrow \rho p$, was presented
in \cite{MRT}.  The cross sections $\sigma_L$ and
$\sigma_T$ in the HERA energy region were
calculated in terms of the known gluon distribution of the
proton.  The approach was based on the \lq open' production of
light quark-antiquark pairs and parton-hadron duality, see
Fig.~1.  That is the diffractive dissociation $\gamma^*
\rightarrow q\overline{q}$ was calculated in a $q\overline{q}$
invariant mass interval around the $\rho$ resonance for high
values of $Q^2$ and high $\gamma^* p$ c.m.\ energy, $W^2 \gg
Q^2$, corresponding to the small $x$ regime.  In the $\rho$ meson
mass interval phase space restrictions
force the $q\overline{q}$ pair to dominately hadronize into two
(or three) pions.  The projections of the $q\overline{q}$ system
into the $I^G = 1^+$ and $0^-$ channels, corresponding to $2\pi$
and $3\pi$ production respectively, are in the ratio 9:1. 
Therefore to obtain the prediction for $\rho$ diffractive
production we multiply the $q\overline{q}$ rate by 0.9.

In order to generalize the approach to the diffractive production
of higher mass $\rho$ states we summarize the relevant formulae.
The cross sections for the production of a $q\overline{q}$ system
of mass $M$ by a photon polarised along, and transverse to, the
$\gamma^*$ direction are respectively given by \cite{MRT}
\bea
\label{eq:a1}
\frac{d\sigma_L}{dM^2} & = & \frac{4 \pi^2 e_q^2
\alpha}{3b} \: \frac{Q^2}{(Q^2 + M^2)^2} \: \frac{1}{8} \:
\int_{-1}^1 \: d \cos \theta \: \left | d_{10}^1 (\theta)
\right |^2 \: \left | I_L \right |^2, \\
& & \nonumber \\
\label{eq:a2}
\frac{d\sigma_T}{dM^2} & = & \frac{4 \pi^2 e_q^2
\alpha}{3b} \: \frac{M^2}{(Q^2 + M^2)^2} \: \frac{1}{4} \:
\int_{-1}^1 \: d \cos \theta \: \left ( \left |d_{11}^1 (\theta)
\right |^2 \: + \: \left |d_{1-1}^1 (\theta) \right |^2 \right )
\: \left | I_T \right |^2,
\eea
where $d_{\lambda \mu}^J (\theta)$ are the usual rotation
matrices.  The outgoing quark has the polar angle $\theta$ in the
$q\overline{q}$ rest frame with respect to the direction of the
incoming proton; that is the transverse momentum of the outgoing
$q$ is $k_T = \frac{1}{2} M \sin \theta$.  The cross sections
\cite{MRT} have been integrated over $t$ assuming an $\exp (-b
|t|)$ behaviour.  We take the observed slope $b = 5.5$
GeV$^{-2}$, consistent with the average experimental value found
in high energy $\rho$ electroproduction for $Q^2 \gapproxeq 10$
GeV$^2$ \cite{RHO}.

The quantities $I_{L,T}$ in (\ref{eq:a1}) and (\ref{eq:a2}) are
integrations over the transverse momenta $\pm \mbox{\boldmath
$\ell$}_T$ of the exchanged gluons \cite{LQ,LMRT,MRT}, see
Fig.~1,
\bea
\label{eq:a3}
I_L (\theta) & = & K^2 \: \int \: \frac{d\ell_T^2}{\ell_T^4} \:
\alpha_S (\ell_T^2) \: f (x, \ell_T^2) \: \left ( \frac{1}{K^2}
\: - \: \frac{1}{K_\ell^2} \right ), \\
& & \nonumber \\
\label{eq:a4}
I_T (\theta) & = & \frac{K^2}{2} \: \int \:
\frac{d\ell_T^2}{\ell_T^4} \: \alpha_S (\ell_T^2) \: f (x,
\ell_T^2) \: \left ( \frac{1}{K^2} \: - \: \frac{1}{2k_T^2} \: +
\: \frac{K^2 - 2k_T^2 + \ell_T^2}{2k_T^2 \: K_\ell^2} \right ),
\eea
where
\bea
\label{eq:a5}
K^2 & = & k_T^2 \: (1 + Q^2/M^2) \; = \; {\textstyle \frac{1}{4}}
\: (Q^2 + M^2) \: \sin^2 \theta, \\
& & \nonumber \\
\label{eq:b5}
K_\ell^2 & \equiv & \sqrt{(K^2 + \ell_T^2)^2 \: - \: 4k_T^2
\ell_T^2},
\eea
and where $f (x, \ell_T^2)$ is the gluon distribution of the
proton, unintegrated over $\ell_T^2$, evaluated at small $x$
given by $x = (Q^2 + M^2)/W^2$.  If we were to assume that the
main contributions to $I_{L,T}$ come from the domain $\ell_T^2 <
K^2$ then we would have the leading $\ln K^2$ approximation
\be
I_L^{LLA} \; = \; I_T^{LLA} \; = \; \frac{\alpha_S (K^2)}{K^2} \:
\int^{K^2} \: \frac{d\ell_T^2}{\ell_T^2} \: f (x, \ell_T^2) \; =
\; \frac{\alpha_S (K^2)}{K^2} \: xg (x, K^2)
\label{eq:a6}
\ee
where $g$, the conventional (integrated) gluon distribution is
sampled at the scale given in (\ref{eq:a5}).  Here we do not use
this approximation, but perform the $d \ell_T^2$ integrations
explicitly.  To be precise, for $\ell_T^2 > \ell_0^2$ we evaluate
the integrals using
\be
f (x, \ell_T^2) \; = \; \frac{\partial (xg (x,
\ell_T^2))}{\partial \ln \ell_T^2}
\label{eq:b6}
\ee
where the gluon distribution $g$ is taken from a recent set of
partons, whereas for the contribution from $\ell_T^2 < \ell_0^2$
we assume that the gluon vanishes linearly with $\ell_T^2$
\be
\alpha_S (\ell_T^2) \: g (x, \ell_T^2) \; = \;
\frac{\ell_T^2}{\ell_0^2} \: \alpha_S (\ell_0^2) \: g (x,
\ell_0^2).
\label{eq:c6}
\ee
This is equivalent to the physical hypothesis that the
gluon-proton cross, $\sigma (gp) \sim \alpha_S g (x,
\ell_T^2)/\ell_T^2$, should tend to a constant (modulo logarithmic 
behaviour) at small values of $\ell_T^2$ and $x$.  We 
choose $\ell_0^2 = 1.75$ GeV$^2$ and test the
sensitivity of the prediction to variation of $\ell_0^2$ about
this value.

Since the perturbative QCD \lq hard' pomeron distorts the initial
$\gamma^* \rightarrow q\overline{q}$ wave function, the
diffractively produced $q\overline{q}$ pair is not
necessarily in a pure $J^P = 1^-$ state.  In fact the
$(q\overline{q})$-proton interaction is proportional to the
square of the separation, $\Delta \rho_T$, of the quarks in the
transverse plane --- an example of the effect of colour
transparency.  The incoming $q\overline{q}$ wave function is
distorted by this $\Delta \rho_T$-dependent amplitude so that the
produced system contains a superposition of the $J^P = 1^-, 3^-,
\ldots \ \ q\overline{q}$ states which are accessible by Pomeron
exchange.  Thus the angular momentum $J$ of
the $q\overline{q}$ state may be changed, although for forward
scattering the $s$ channel helicity is still conserved.  The
distortion enters the cross section formula (\ref{eq:a1}) and
(\ref{eq:a2}) through the $\theta$ dependence of $I_{L,T}$; the
integrals which arise from the $q\overline{q}$ interaction with
the proton.  The $q\overline{q}$ production amplitudes which
enter (\ref{eq:a1}) and (\ref{eq:a2}) may therefore be decomposed
in the form
\be
d_{1 \lambda}^1 (\theta) \: I_i (\theta) \; = \; \sum_J \: c_i^J
(\lambda) \: d_{1 \lambda}^J (\theta),
\label{eq:d6}
\ee
where $i = L$ or $T$, so that the probability amplitude for spin
$J$ production is proportional to the coefficient
\be
c_i^J (\lambda) \; = \; \frac{2J + 1}{2} \: \int \: d \cos \theta
\: \left [ d_{1 \lambda}^1 (\theta) \: I_i (\theta) \right ] \:
d_{1 \lambda}^J (\theta)
\label{eq:e6}
\ee
and so spin $J$ production in longitudinally and transversely
polarised states is proportional to $|c_L^J (0)|^2$ and $|c_T^J
(1)|^2 + |c_T^J (- 1)|^2$ respectively.  Clearly if the $I_i
(\theta)$ were independent of $\theta$ then
$J^P = 3^-$ $q\overline{q}$ production would be zero, and pure $J^P =
1^-$ production would occur.  The integrals are infrared
convergent not only for $\sigma_L$, but also for $\sigma_T$ ---
we quantify the infrared sensitivity of the predictions below.

In ref.\ \cite{MRT} parton-hadron duality was invoked to estimate
diffractive $\rho$ electroproduction.  The $q\overline{q}$
production cross section, projected into the $1^-$ channel, was
integrated over the mass interval $0.6 < M < 1.05$ GeV.  We
summed over $u\overline{u}$ and $d\overline{d}$ production and
multiplied by 0.9 to allow for $\omega$ production.  We found
good agreement with the measurements of $\gamma^* p \rightarrow
\rho p$ obtained at HERA \cite{RHO}.  Of course the absolute
normalisation of the cross section depends on the choice of mass
interval and also on the scale of $\alpha_S$ used to estimate the
$K$ factor enhancement coming from higher order contributions
(which is evaluated as described in refs.\ \cite{LMRT,MRT}). 
Thus it is the $Q^2$ behaviour of the total cross section and,
more particularly, of the ratio $\sigma_L/\sigma_T$ which offer
tests of the model.  To the best of our knowledge, this is the
first time that the $Q^2$ behaviour of the ratio
$\sigma_L/\sigma_T$ has been satisfactorily described.

Now measurements of diffractive electroproduction into higher
mass pionic systems are becoming available, for instance
$\gamma^* \rightarrow 4 \pi$ has been observed in the mass range
$1.4 < M < 2$ GeV \cite{RHO}.  Indeed there is evidence of a
broad peak for $M \sim 1.6$ GeV in the $4 \pi$ channel, which may
be attributed to a combination of the $\rho^\prime (1450, 1700)$
and $\rho (1690)$ resonances with $J^P = 1^-$ and $3^-$
respectively.  To provide estimates of the QCD expectations for
the diffractive production of these resonances we again invoke
parton-hadron duality.\footnote{An alternative approach for
$\rho^\prime$ production, which is based on the colour dipole model,
can be found in \cite{NNN}.}  In this case we integrate the $J = 1$ and
$J = 3$ projections over the mass interval $1.3 < M < 1.8$ GeV. 
Again we multiply by 0.9 to project onto the $I = 1$ channel. 
The results are shown in Figs.\ 2 and 3 by the curves labelled
$\rho^\prime (1^-)$ and $\rho (3^-)$ respectively.  We also show the
results for $\rho$ meson production for comparison.  The gluon of
the MRS(R2) set of partons \cite{MRSR} is used.

\begin{figure}[htb]
\begin{center}
%\vskip -3mm
\leavevmode
\epsfxsize=17.cm
\epsffile[70 240 490 570]{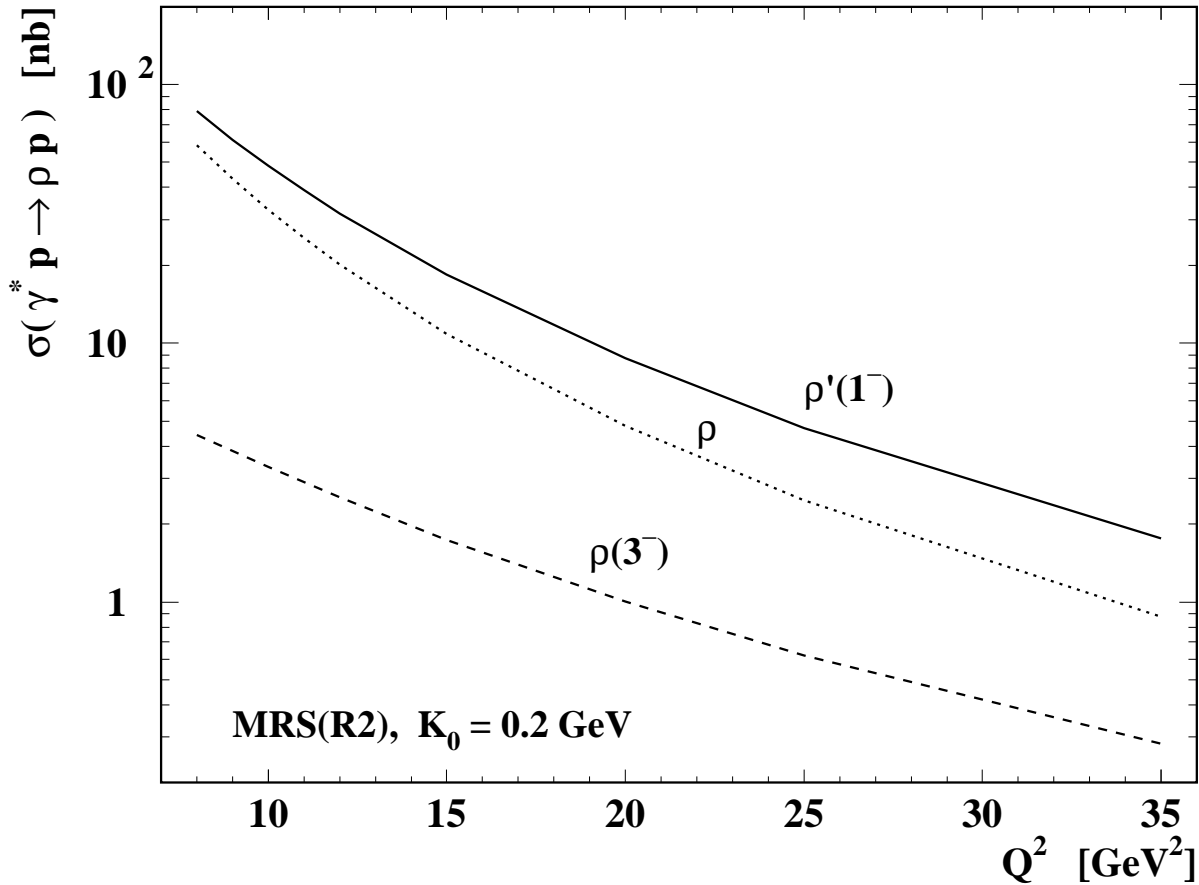}
\vskip -5mm
\caption[]{\label{fig2} {The QCD predictions for the total cross section for
diffractive $\rho (1^-, 770\ {\rm MeV}),\ \rho^\prime (1^-)$ and 
$\rho (3^-)$ electroproduction as a function of the photon
virtuality $Q^2$, where the gluon distribution $g(x, l_T^2)$ is
evaluated at $x = (Q^2 + M^2)/W^2$ with $W = 83.6$~GeV.}} 
\end{center}
\end{figure}
\begin{figure}[htb]
\begin{center}
%\vskip -3mm
\leavevmode
\epsfxsize=17.cm
\epsffile[70 240 490 570]{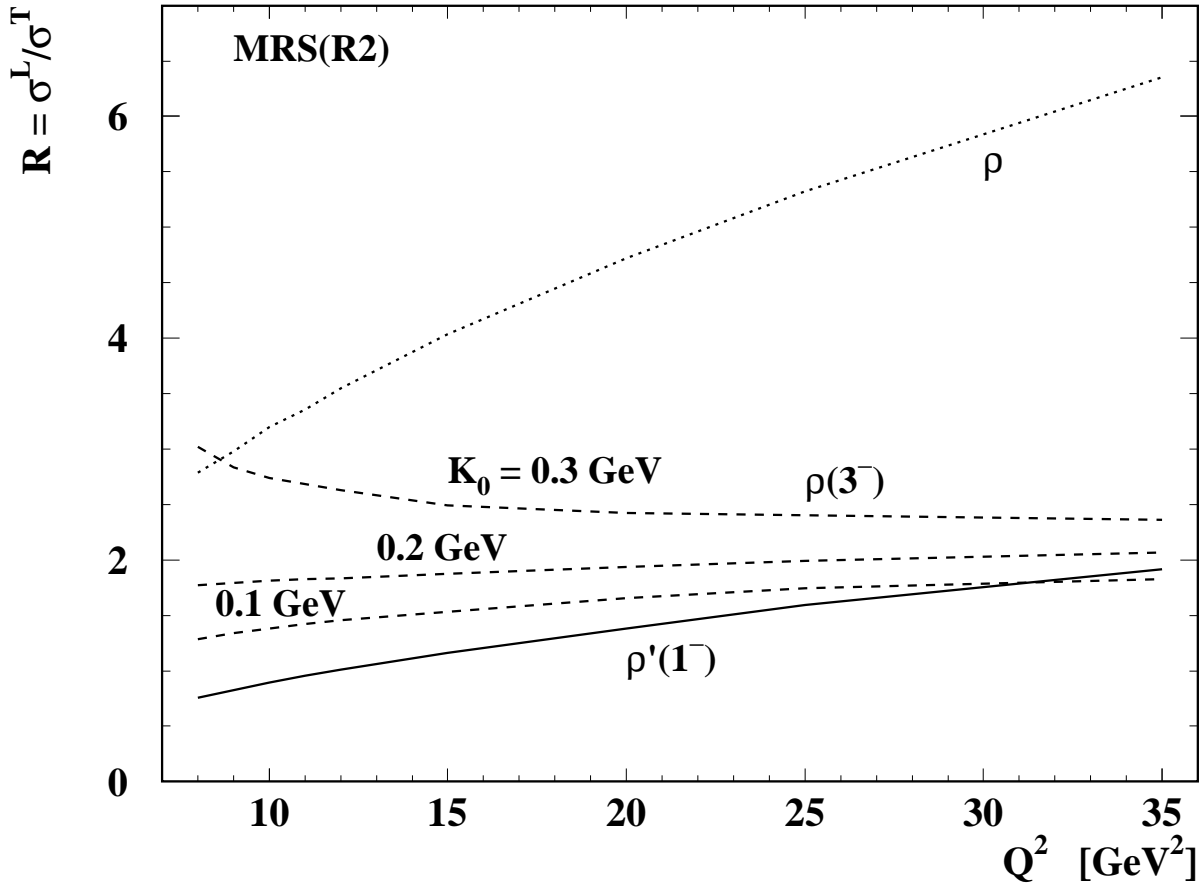}
\vskip -5mm
\caption[]{\label{fig3} {The QCD predictions for $R = \sigma_L/\sigma_T$ as
a function of $Q^2$ for $\rho (1^-, 770\ {\rm MeV}),\ \rho^\prime
(1^-)$ and $\rho (3^-)$ diffractive electroproduction 
for $W = 83.6$ GeV.  For $3^-$
production we show predictions for three different choices of the
infrared cut-off $K_0$.}} 
\end{center}
\end{figure}

How dependent are the predictions on the treatment of the
infrared regions of low transverse momenta of the exchanged
gluons $\pm \mbox{\boldmath $\ell$}_T$ and of the produced quarks
$\pm \mbox{\boldmath $k$}_T$?  First we note that the results are
hardly sensitive to variation of the value of $\ell_0^2$ below
which we assume that the gluon distribution vanishes linearly in
$\ell_T^2$, see (\ref{eq:c6}).  For convenience we take $\ell_0^2
= 1.75$ GeV$^2$, but essentially no change occurs for different
choices of $\ell_0^2$ about this value.

We now come to the integration over the quark transverse momentum
$k_T$, which in (\ref{eq:e6}) has been transformed into an integration
over $\cos \theta$. In principle, in pure perturbative QCD we should
integrate essentially down to $k_T = 0$.  However, confinement
eliminates the large distance contribution and hence we impose a
cut-off\footnote{The essential distances in the process are
controlled by $K$, which is related to $k_T$ by eq.\
(\ref{eq:a5}) \cite{LQ,LMRT}.} $K_0 \sim 1$ fm$^{-1} 
\simeq 0.2$~GeV.  The results shown in Figs.\ 2 and 3 are obtained
with $K_0 = 0.2$~GeV.  That is we cut-off the
low $\sin \theta$ contributions from the projection integrals
given in eq.\ (\ref{eq:e6}).  The sensitivity to the choice of
$K_0$ is displayed in Table 1, which shows the diffractive
electroproduction cross sections obtained by taking first $K_0 =
0.1$~GeV and, second, $K_0 = 0.3$~GeV.  It is evident from Fig.~3
that the sensitivity to the cut-off becomes less for large $Q^2$,
since the infrared domain $K^2 < K_0^2$ corresponds to an
increasingly smaller part of the region of integration, see
(\ref{eq:a5}).  By far the
biggest uncertainty is in the prediction of $\sigma_T$ for $3^-$
production.  This is to be expected.  For $\sigma_L$ the
contribution from the infrared region of small $\sin \theta$ in
(\ref{eq:e6}) is suppressed since $d_{10}^1 d_{10}^J \sim \sin^2
\theta$ as $\sin \theta \rightarrow 0$ for both $J = 1$ and 3. 
However, the suppression factor is absent in $\sigma_T$.  Now the
distortion $I_i (\theta)$ of the $q\overline{q}$ state behaves
approximately as $(\sin^2 \theta)^{\gamma - 1}$ where $\gamma$ is
the anomalous dimension of the gluon, $g (x, K^2) \sim
(K^2)^\gamma$.  Thus for $\sigma_T$ the distortion is larger in
the region of small $\sin \theta$.  We note that the projection
onto the higher $J$ state is more sensitive to this region, due
to the behaviour of $d_{1, \pm 1}^J$ for small $\sin \theta$.  A
comparison of the
predictions obtained with $K_0 = 0.1$ and 0.3 GeV shows just
these trends.  $\sigma_L$ is much more insensitive to variation
of $K_0$ than is $\sigma_T$, and $1^-$ production is less
sensitive than $3^-$ production.

\begin{table}[htb]
\begin{center}
\begin{tabular}{|c|cc|cc|} \hline
& \multicolumn{2}{|c|}{$\sigma_L (nb)$} &
\multicolumn{2}{|c|}{$\sigma_T (nb)$} \\ \hline
$K_0$ (MeV) & 100 & 300 & 100 & 300 \\ \hline
$\rho$ & 25.0 & 24.8 & 8.2 & 7.2 \\
$\rho^\prime(1^-)$ & 22.8 & 22.8 & 26.7 & 24.0 \\
$\rho (3^-)$ & 2.2 & 2.1 & 1.6 & 0.8 \\ \hline
\end{tabular}
\end{center}
\vskip -1mm
\caption{The cross sections $\sigma_{L, T} (\gamma^* p \rightarrow \rho
p)$ for diffractive electroproduction of longitudinally and
transversely polarised $\rho$ resonances, for $Q^2 = 10$ GeV$^2$
for two different choices of the infrared cut-off $K_0$ of the
variable $K$ defined by (5). The $\rho$ cross sections come from
integrating $q\overline{q}$ production over the mass interval
$0.6 < M < 1.05$ GeV, whereas $\rho^\prime (1^-)$ and $\rho
(3^-)$ correspond to integration over the interval $1.3 < M <
1.8$ GeV.}
\end{table}

From Fig.\ 2 and Table 1 we see that $\rho^\prime (1^-)$
diffractive electroproduction is predicted to occur at a
comparable rate to $\rho$ production at the lower values of $Q^2$
shown, but decreases more slowly as $Q^2$ increases.  Fig.~3
shows that the expected values of $R = \sigma_L/\sigma_T$ are
smaller for the higher mass states than those for $\rho$
production.  The main reason is the presence of the kinematic
factor $Q^2/M^2$ in $R$.  We also see that $3^-$ production is
not insignificant.  The $\rho (3^-)$ state is expected to occur
at about 0.1 the rate of $\rho^\prime$ production.  The cross
section $\sigma_L (3^-)$ is much better determined than $\sigma_T
(3^-)$; the latter has about a factor of 2 uncertainty. 

We conclude that if a sufficient number of diffractive
electroproduction events are observed at HERA so as to be able to
separate the $1^-$ and $3^-$ systems, then these processes will
allow new insights of the \lq QCD' pomeron.  However, first
we must consider the hadronization of the produced
$q\overline{q}$ pair.  Recall that in the $\rho (770)$ mass
region, phase space restrictions force the $q\overline{q}$
pair to hadronize dominantly into two pions.  In the higher mass
region the situation is more complicated.  From the branching
ratios of the resonances listed in the PDG tables \cite{PDG} we
expect that the $q\overline{q}$ state hadronizes into the $4 \pi$
channel some 70--80\% of the time, with about 20\% going into $2
\pi$.  Of course it is not easy to perform a partial wave
analysis of the $4 \pi$ channel to separate out the $3^-$
component.  Fortunately the $\rho (1690)$ has healthy branching
ratios into two hadron decay modes; for instance ${\rm BR}
(\rho^0 \rightarrow \omega \pi^0) = 16\%$ and ${\rm BR} (\rho
\rightarrow 2 \pi) = 24\%$.

It would also be very interesting to observe the diffractive
$\gamma^* \rightarrow q\overline{q}$ dissociation into the $2^+$
resonant state, that is $\gamma^* \rightarrow f_2 (1270)$. 
Unfortunately the cross section is very small.  There are two
main reasons for the suppression.  First, the resonance isospin
$I = 0$ gives rise to a 1:9 suppression in comparison to $I =
1~\rho$ electroproduction, due to the electric charges of the
quarks.  Second, the $\gamma^* (1^{- \: -}) \rightarrow f_2 (2^{+
\: +})$ production process occurs via odderon, rather than
pomeron, exchange.  The odderon amplitude is mainly real and in
perturbative QCD is given by 3 gluon exchange, to leading order. 
Numerical estimates \cite{M} indicate that $|A ({\rm
odderon})|/|A (\funp)| \lapproxeq 0.03$, which agrees well with
the limits obtained from the ratio of the real to the imaginary
part of the forward scattering amplitude measured in $pp$ and
$p\overline{p}$ elastic scattering \cite{PP}.  So we expect a
suppression of $2^+ \ q\overline{q}$ production of at least
$\frac{1}{9} (0.03)^2 \sim 10^{-4}$ in comparison to diffractive
$\rho$ production.  We thus estimate a $\gamma^* \rightarrow f_2$
cross section will be less than 1 pb at $Q^2 = 10$ GeV$^2$ which
will make observation at HERA very difficult.

%\newpage

\end{document}